\documentclass[sigplan,screen]{acmart}

\setcopyright{acmlicensed}
\copyrightyear{2018}
\acmYear{2018}
\acmDOI{XXXXXXX.XXXXXXX}

\acmConference[Conference acronym 'XX]{Make sure to enter the correct
  conference title from your rights confirmation emai}{June 03--05,
  2018}{Woodstock, NY}
\acmISBN{978-1-4503-XXXX-X/18/06}

\usepackage{amsmath}
\usepackage{caption}
\usepackage[shortcuts]{extdash} 
\usepackage{mathtools}
\usepackage{mdframed}
\usepackage[frozencache=true,cachedir=minted-cache]{minted}
\usepackage{newunicodechar}
\usepackage{phoenician}
\usepackage{textgreek}
\usepackage{xspace}
\usepackage{enumitem}

\usepackage [autostyle, english = american]{csquotes}

\newunicodechar{λ}{\ifmmode\lambda\else\textlambda\fi}

\MakeOuterQuote{"}

\newcommand{\HLS}[1][small]{$\rotatebox[origin=c]{15}{\textphnc{e}}\hspace{-1pt}_{λ\mathrm{#1}}$\xspace}
\newcommand{\ie}{\textit{i.e.}\xspace}
\newcommand{\HasChor}{Has\-Chor\xspace}
\newcommand{\inlinecode}[2][haskell]{\mintinline[breaklines]{#1}{#2}}
\newcommand{\MultiChor}{\texttt{Multi\-Chor}\xspace}

\begin{document}

\title[MultiChor]{MultiChor: Census Polymorphic Choreographic Programming with Multiply Located Values}

\author{Mako Bates}
\email{mako.bates@uvm.edu}
\orcid{0009-0001-9933-1728}
\affiliation{%
 \institution{University of Vermont}
 \city{Burlington}
 \state{Vermont}
 \country{US}}

\author{Joseph P. Near}
\email{jnear@uvm.edu}
\orcid{0000-0002-3203-3742}
\affiliation{%
 \institution{University of Vermont}
 \city{Burlington}
 \state{Vermont}
 \country{US}}
 
\author{Syed Jafri}
\email{sajafri@uvm.edu}
\orcid{0009-0009-4372-4715}
\affiliation{%
 \institution{University of Vermont}
 \city{Burlington}
 \state{Vermont}
 \country{US}}

\renewcommand{\shortauthors}{Trovato et al.}

\begin{abstract}
  Choreographic programming is a concurrent paradigm
  in which a single global program called a choreography
  describes behavior across an entire distributed network of participants.
  Choreographies are easier to reason about than separate programs running in parallel,
  and choreographic programming systems can check for deadlocks statically.

  We present \MultiChor, a library for writing and running choreographies as
  monadic values in Haskell.
  Unlike prior Haskell implementations,
  \MultiChor does not require excess communication to handle Knowledge-of-Choice.
  Unlike all prior general-purpose choreographic languages,
  \MultiChor can express choreographies that are polymorphic over the number of participants.
\end{abstract}



\received{20 February 2007}
\received[revised]{12 March 2009}
\received[accepted]{5 June 2009}

\maketitle

\section{Introduction}\label{sec:introduction}
Choreographic programming languages facilitate concurrent programming
both by offering excellent ergonomics to programmers,
and by statically ensuring deadlock freedom and preventing other errors
specific to concurrent settings.
Despite decades of evolving and advancing theory,
implementations of choreographic programming \textit{per se} are relatively young.

One limitation on the use of choreographies has been that they encode exactly who is doing what,
while protocols used in the real world are often
parameterized by their participants.
Location-polymorphism~\cite{graversen2023polychor} addressed part of this limitation
by letting roles in a choreography be filled by different participants in different cases,
but did not enable a choreography to work for different \emph{numbers of} participants.
This lack of polymorphism over the number of participants in previous approaches makes it impossible to write general implementations of many important classes of practical protocols, including approaches for federated machine learning~\cite{bonawitz2019towards, wu2020safa, li2021model} and secure computation~\cite{keller2020mp, corrigan2017prio, bonawitz2017practical, dprio2023, goldreich2019play}.

Choreographic programming was introduced to the Haskell ecosystem
in 2023 in the form of the \HasChor library~\cite{haschor};
\HasChor's techniques for \emph{implementing} choreographic systems have proven effective
and adaptable (\textit{e.g.} the comparable implementation of ChoRus~\cite{chorus}),
but the communication efficiency of these approaches can sometimes be worse than previous standalone systems (e.g. Choral~\cite{choral}).

We present \MultiChor, a new library for writing choreographies in Haskell. \MultiChor enables efficient communication via \emph{multiply located values} and supports \emph{census polymorphism} to allow variable numbers of participants. Our case studies demonstrate the use of \MultiChor to implement general versions of variable-participant protocols, including two for secure computation.

\paragraph{Contributions}\label{sec:contributions}
In summary, our contributions are:
\begin{itemize}[leftmargin=12pt]
    \item A working implementation of the multi\-/local\-/\&\-/multicast choreographic paradigm
    in Haskell.
    Compared to prior implementations, this system allows better communication efficiency
    and concise expression of parallel computation.
    \item Novel features for a choreographic system:
    census polymorphism, \inlinecode{fanOut}, and \inlinecode{fanIn}.
    These allow \MultiChor to express complicated protocols with variable numbers of participants.
    \item Examples demonstrating how these language features, and \MultiChor specifically,
    can concisely and correctly implement complex choreographies,
    including the GMW protocol for secure computation~\cite{goldreich2019play}.
\end{itemize}

\section{Choreographic Programming}\label{sec:background}

Choreographic programming is a language paradigm for concurrent programming in which
the actions and computations of all participants (locations, parties, or endpoints)
are described in a single unified program called a choreography that enables global reasoning about the semantics of the distributed system.
A fundamental attribute of choreographic languages is that
sending and receiving data are not separate operations;
these "two sides of the same coin" are represented as a single action or effect.
For example, the \MultiChor expression
\inlinecode{(alice, invitation) ~> bob}
is interpreted by \inlinecode{alice} as \textit{"send an invitation to \inlinecode{bob}"},
and by \inlinecode{bob} as \textit{"receive an invitation from \inlinecode{alice}"}.
Figure~\ref{fig:card-game} shows a more involved example;
a card game with a dealer and many players is described as a single Haskell expression
in the \inlinecode{Choreo} monad.
Choreographic languages define \emph{endpoint projection} (EPP),
a compilation step in which an individual party's behavior is extracted from a choreography.
For example, a player's EPP of the \inlinecode{fanOut} loop on
line~17 of Figure~\ref{fig:card-game}
would ignore all but one of the loop iterations
because of the \inlinecode{enclaveTo} on line~18.
The dealer's EPP of the \inlinecode{fanOut} would have interaction with a player
in every iteration.

\begin{figure*}[tbhp]
    \begin{mdframed}
\begin{minted}[xleftmargin=10pt,linenos,fontsize=\small]{haskell}
{- A simple black-jack-style game. The dealer gives everyone a card, face up. Each player may
 - request a second card. Then the dealer reveals one more card that applies to everyone. Each
 - player individually wins if the sum of their cards (modulo 21) is greater than 19.  -}
game :: forall players m. (KnownSymbols players) => Choreo ("dealer"': players) (CLI m) ()
game = do
  let players = consSuper (allOf @players)
      dealer = listedFirst @"dealer"
  hand1 <- fanOut players \player -> do
      card1 <- dealer `_locally` getInput ("Enter random card for " ++ toLocTm player)
      (dealer, card1) ~> inSuper players player @@ nobody
  onTheTable <- fanIn players players \player -> do
      (player, players, hand1) ~> players
  wantsNextCard <- players `parallel` \player un -> do
      putNote $ "My first card is: " ++ show (un player hand1)
      putNote $ "Cards on the table: " ++ show (un player onTheTable)
      getInput "I'll ask for another? [True/False]"
  hand2 <- fanOut players \(player :: Member player players) -> do
      (dealer @@ inSuper players player @@ nobody `enclaveTo` listedSecond @@ nobody) do
        choice <- broadcast (listedSecond @player, (player, wantsNextCard))
        if choice then do
            cd2 <- dealer `_locally` getInput (toLocTm player ++ "'s second card:")
            card2 <- (dealer, cd2) ~> listedSecond @@ nobody
            listedSecond `locally` \un -> pure [un player hand1, un singleton card2]
          else listedSecond `locally` \un -> pure [un player hand1]
  tblCrd <- dealer `_locally` getInput "Enter a single card for everyone:"
  tableCard <- (dealer, tblCrd) ~> players
  players `parallel_` \player un -> do
      let hand = un player tableCard : un player hand2
      putNote $ "My hand: " ++ show hand
      putOutput "My win result:" $ sum hand > card 19
\end{minted}
    \caption{A card game expressed as a choreography written in \MultiChor.
             This choreography is polymorphic over the number and identity of the players,
             but the party named \inlinecode{"dealer"} is an explicit member.
             The inner monad \inlinecode{CLI} that all parties have access to is a simple freer monad
             that can be handled to IO operations, or as \inlinecode{State} for testing purposes.
             The \inlinecode{newtype Card} encapsulates the modulo operation in its
             \inlinecode{Num} instance.}
    \label{fig:card-game}
    \Description{Thirty lines of haskell code describing a choreography called "game".}
    \end{mdframed}
\end{figure*}

\paragraph{Limitations of prior work}
Our work aims to fill two gaps in existing work on choreographic programming:
\begin{enumerate}[leftmargin=12pt, itemsep=0pt]
\item Prior library-based implementations of choreographic programming sometimes add unnecessary communication.
\item Prior work does not support \emph{census polymorphism} --- choreographies parameterized by both the quantity and the identities of participants.
\end{enumerate}
The rest of this section describes related work in the context of these limitations; Section~\ref{sec:library} introduces \MultiChor, our library-based approach for choreographic programming that addresses both limitations.

\paragraph{Census}\label{sec:census}
It is common for a choreography to be associated with the list of parties who participate in it;
we call such a list a "\textbf{census}".
When considering a choreography whose census is known, one may \emph{safely pretend} that
the census is an exhaustive list of all parties in existence.
(We do not consider it problematic if a party is listed in a census but 
doesn't actually do anything in the respective choreography.)
Chorλ~\cite{chor-lambda}, Pirouette~\cite{hirsch2021pirouette}, and \HLS~\cite{bates2024know}
track censuses as context in their type systems
but \HasChor~\cite{haschor} operates without any analogous concept at all.
In \MultiChor the first type argument to \inlinecode{Choreo} is the census.

We define \emph{census polymorphism} to be polymorphism over both the size and contents of a choreography's census. Many prior approaches support polymorphism over the \emph{identities} of census members,
but not over its size~\cite{chor-lambda, graversen2023polychor, hirsch2021pirouette, bates2024know, haschor}.

\paragraph{Knowledge of Choice}\label{sec:koc}
Part of what makes choreographic programming non-trivial is that every time any party
makes a choice in control-flow (\ie any time a party "branches"),
other parties may need to know which branch was taken in order to correctly participate in the rest of the choreography.
This is called "Knowledge of Choice" or "KoC".
A common KoC strategy in theoretical choreography research is to use explicit \inlinecode[text]{select}
statements to inform parties, as needed, that a branch has been chosen.
\inlinecode[text]{select} is typically paired with a "merge" process during EPP,
which allows parties whose role across different branches is invariant to participate \emph{without}
KoC.
The select-\&-merge paradigm offers optimal communication efficiency,
but is difficult to implement as a type-safe library because it requires a custom static analysis during EPP. As a result, existing implementations of this approach for KoC have been via standalone compilers, rather than as libraries for existing mainstream programming languages.

\paragraph{HasChor}\label{sec:haschor}
Shen~\textit{et~al}~\cite{haschor} presented \HasChor,
a Haskell library for writing and running choreographies.
Unlike most earlier systems, \HasChor is "just a library",
an embedded DSL instead of standalone language like \HLS and Chorλ
or a layered system like Choral~\cite{choral} and Pirouette~\cite{hirsch2021pirouette};
This is a significant advantage for industry use;
getting started using \HasChor is as simple as adding it as a dependency
in one's \inlinecode{cabal} file,
and \HasChor's \inlinecode{Choreo} monad can be used with the copious tools
Haskell offers for monadic programming.
On the other hand, \HasChor's KoC strategy was always understood to be inefficient;
the guard of every conditional is broadcast to everyone
regardless of who actually needs to know it.
Furthermore, \HasChor doesn't track any kind of census as a property of the choreographies;
the broadcasts simply go to every location the branching process knows how to contact.

\paragraph{ChoRus}
Kashiwa~\textit{et~al}~\cite{chorus} introduces ChoRus,
a Rust library offering similar choreographic programming to \HasChor,
but with explicit census tracking.
ChoRus introduces an "enclave" operator to avoid the superfluous communication of \HasChor,
but Bates \& Near~\cite{bates2024know} show that there are situations where it can not achieve
state-of-the-art efficiency.

\paragraph{He-Lambda-Small}\label{sec:he-lambda-small}
Bates~\& Near~\cite{bates2024know} presented \HLS, a choreographic lambda calculus
whose KoC strategy relies on \emph{multiply located values}.
A multiply located value is a single value that's known to multiple parties.
\HLS uses case-expressions for branching, and restricts the census inside the case expressions
to only the parties who know the guard value.
Multiply-located values pair well with a multicast operator;
instead of the primitive \inlinecode{com} operation representing a message
from one party to another,
in \HLS \inlinecode{com} represents a message from a single party to a \emph{set} of parties.
Bates~\&~Near show that in theory this can match the communication efficiency of
state-of-the-art select-\&-merge languages like Chorλ~\cite{chor-lambda},
and show by several examples that multi-local-\&-multicast choreographic programming
is expressive and ergonomic. 
Additionally, the validity of a \HLS choreography is entirely type-directed and EPP is guaranteed to succeed for well-typed choreographies.
\HLS's multiply-located values thus offer a third alternative to the select-\&-merge paradigm (which is not amenable to library-based implementation) and \HasChor's broadcast paradigm (which does not offer optimal communication).

\paragraph{Other related work}\label{sec:related-work}
Wysteria~\cite{wysteria} and λ-Symphony~\cite{Sweet_2023} are purpose-specific languages
for working with particular kinds of multi-party cryptography.
Programs in these languages are choreographies, and can exhibit census polymorphism,
but both of these languages have homomorphic encryption baked into their
semantics for communication,
and they cannot be used for general-purpose choreographic programming.
Jongmans~\&~van~den~Bos~\cite{jongmans2022predicates}
present a KoC strategy that subsumes \HLS's KoC strategy;
their system does not use multiply located values, and instead uses predicate transformers
on the semantics of programs to check that distributed decisions are unanimous.
Pirouette~\cite{hirsch2021pirouette}, Chorλ~\cite{chor-lambda},
and  PolyChorλ~\cite{graversen2023polychor}
are functional languages for writing select-\&-merge choreographies;
PolChorλ introduces polymorphism over identities of parties.
Choral~\cite{choral} is a choreographic language implementing the select-\&-merge paradigm
and targeting industrial use;
it runs on the JVM and can easily import local Java code.

\section{The MultiChor library}\label{sec:library}

We present a new "just a library" choreographic programming system for Haskell: \MultiChor.

Our work adapts ideas from \HLS to address both limitations of prior work described in Section~\ref{sec:background}. With \MultiChor, we show that multiply-located values simultaneously enable a library-based implementation and optimal communication, addressing limitation (1). In addition, by representing the census as a type-level variable in Haskell, \MultiChor enables polymorphism over both the size and membership of the census, addressing limitation (2).

\subsection{Desiderata}\label{sec:wants}

The first goal which \MultiChor accomplishes is to combine the accessibility of \HasChor
with the KoC strategy, communication efficiency, and ergonomics, of \HLS.

A key innovation of \HLS is that KoC is enforced entirely
by type-level management of the census.
For implementation in Haskell, it's natural to represent the census as a type-level-list
argument to the type of choreographies.
Immediately a handful of attributes occur as desirable for such a system:
\begin{enumerate}
    \item Choreographies should have censuses that are statically enforced by the type system.
    \item It should be possible to write a choreography that's polymorphic over its census.
    \item It should be possible to broadcast, \ie to multicast a value to the entire census,
          and to use values known to the entire census as normal (un-located) values of their type.
    \item It should be possible to know from an appropriately-written choreography's type that some          certain party or parties are not involved, are not in its census.
          Users should be able to embed such "enclave" choreographies inside choreographies with larger,
          possibly polymorphic, censuses.
    \item The type system should be able to reason about parties'
          membership in a census or ownership-set
          with normal subset reasoning.
\end{enumerate}
The choreography in Figure~\ref{fig:card-game} showcases all of the above points.
The census of the whole program appears in the type (\#1)
and does not specify who the players are (\#2).
The \inlinecode{enclaveTo} on line~18
embeds a choreography whose census is exactly the monomorphic \inlinecode{"dealer"}
and a polymorphic \inlinecode{player} (\#4).
The helper-function \inlinecode{broadcast} on line 19 functions as described in \#3.
Many examples of \#5 are automated or hidden in \MultiChor,
but on line~19 the function \inlinecode{inSuper} is applied to
\inlinecode{players :: Subset players ("dealer" : players)}
and \inlinecode{player :: Member player players}
to attest that \inlinecode{player} is present in the census.

\paragraph{Reasoning about location-sets}
To illuminate point 5 above, consider the program in Figure~\ref{fig:example-transitive},
a choreography with a polymorphic census \inlinecode{census}.
For this to be a well-formed choreography, all of the following must be true
(to support the parenthesized statements):
\begin{itemize}
    \item \inlinecode{clique} must be a subset of \inlinecode{census}.
          (\inlinecode{enclave} on line~3)
    \item \inlinecode{bob} must be a member of \inlinecode{census}.
          (\inlinecode{~>} ("send") on line~4)
    \item \inlinecode{bob} must be a member of \inlinecode{clique}.
          (\inlinecode{~>} on line~6, as well as \inlinecode{~>} on line~4 so they can use \inlinecode{theirFoo})
\end{itemize}
Notice that one of these prerequisites is redundant:
if Bob is a member of \inlinecode{clique}, and \inlinecode{clique} is a subset of \inlinecode{census},
then \emph{it follows} that Bob is a member of \inlinecode{census}.
When all the sets of parties are explicitly listed out, memberships and subsets can simply be observed,
but reasoning about polymorphic sets is difficult for Haskell's type system.
Leveraging the transitivity of subset relations is especially difficult~\cite{stackoverflow2021},
and without such reasoning it's difficult to write reuseable code.
We overcome this challenge using Ghosts of Departed Proofs~\cite{noonanGDP};
the proof-objects do double-duty in \MultiChor as both proofs that operations are legal \emph{and}
as \inlinecode{Proxy}-like identifiers for type-level parties and lists of parties.

\begin{figure*}[tbhp]
    \begin{mdframed}
\begin{minted}[xleftmargin=10pt,linenos,fontsize=\small]{haskell}
exampleChor :: Choreo census m (Located '["carroll"] Int)  -- The inner monad m is not used here.
exampleChor = do
    theirFoo <- clique `enclave` foo  -- Lift a Choreo of "clique" into a Choreo of "census".
    (bob, theirFoo) ~> carroll @@ nobody  -- Bob sends `theirFoo` to Carroll (and nobody else).
  where foo = do
          aliceFoo <- (bob, bobFoo) ~> alice @@ nobody  -- Bob sends his value `bobFoo` to Alice.
          broadcast (alice, aliceFoo)  -- Alice sends it to the census of `foo`,
                                       -- _not_ the census of `exampleChor`.
\end{minted}
    \caption{An example choreography, written with \MultiChor, in which an \inlinecode{Int},
             originating at a party \inlinecode{bob}, is passed back and forth.
             Ultimately, a version of that value owned only by \inlinecode{carroll} is returned.
             Part of the choreography takes place in an \inlinecode{enclave}
             involving some collection of parties specified by \inlinecode{clique}.}
    \label{fig:example-transitive}
    \Description{Eight lines of haskell code describing a choreography called "exampleChor".}
    \end{mdframed}
\end{figure*}

\paragraph{Congruent computation}
One of the attractive features of multiply-located values is that they allow concise expression of
congruent computations.
A \emph{congruent} computation is one that is performed in parallel by multiple parties
deterministically on \emph{the same} values, such that all parties arrive at the same result
(\textit{e.g.} Figure~\ref{fig:lottery} line~37).
This is distinct from a \emph{parallel} computation
(\textit{e.g.} Figure~\ref{fig:lottery} line~26), whose distributed versions are expressed
by the same possibly-branching algorithm but might be operating on different data
and arrive at different results.
Parallel computation is more heavily studied and more widely used;
congruent computation is desirable when communication is more expensive (slower)
than local computation.
We believe a choreography library should support both congruent and parallel computation
as part of type-safe use of located values.

\subsection{Location-set and census polymorphism}\label{sec:located-faceted}

Being a proof-of-concept lambda calculus, \HLS doesn't support polymorphism of data-types
or of locations.
\HasChor has a blunter API for writing choreographies, but as an eDSL it can apply
Haskell's polymorphism "for free".
\MultiChor is a similar eDSL, with some of \HasChor's skeleton intact at its heart,
and can similarly apply Haskell's polymorphism to its \emph{lists} of parties.
(In addition to polymorphism over a census, \MultiChor expressions
can also be polymorphic over lists of data-owners;
it's rarely important to distinguish between these features.)
Without complementary features this would be
unsatisfying---the only way an unidentified and variable mass of parties in a census could actually
participate is as recipients of broadcasts.

The first step toward \emph{useful} location-set polymorphism is a data type analogous to
multiply-located values (\inlinecode{Located (ps::[LocTy]) a} in \MultiChor)
but without the guarantee that the parties' respective values will all the the same.
We call such structures \inlinecode{Faceted}, after the many facets that form
the surface of a cut gem-stone.
This language feature immediately enables an operation similar to \HasChor's \inlinecode{locally},
except that \MultiChor's \inlinecode{parallel} is evaluated by a list of locations.
The block in Figure~\ref{fig:card-game} lines~13--16 has type 
\inlinecode{Choreo ... (Faceted players Bool)}
and yields a single "value" that \emph{represents} the likely-different booleans read from
\inlinecode[text]{stdin} by the (polymorphic) parties \inlinecode{players}.
\inlinecode{Faceted} values can be multicast by their owners
and used in future \inlinecode{parallel} blocks
just like \inlinecode{Located} values.

\paragraph{Fan-Out and Fan-In}
\inlinecode{Faceted} values are still insufficient for two tasks we believe a location-set polymorphic
choreography library should support:
\begin{enumerate}[leftmargin=12pt, itemsep=0pt]
\item A party should be able to send each member of a polymorphic list a \emph{distinct} value.
\item The members of a polymorphic list should be able to \emph{send} messages.
\end{enumerate}
We accomplish these with two subtly different primitives:
\inlinecode{fanOut} and \inlinecode{fanIn}.
Both take as an argument a choreography with a single-location parameter,
and instantiate the choreography with each member of the specified location set,
like a \inlinecode[text]{foreach} loop.
In the case of \inlinecode{fanOut}, the loop-body is required to return a value at the subject 
location;
these values get packaged together as a \inlinecode{Faceted}.
For \inlinecode{fanIn}, the return value must be located at a \emph{consistent} set of recipients;
the values get packaged as a \inlinecode{Located recipients [value]}.
Both of these operations are used in Figure~\ref{fig:card-game}.

\subsection{The Core API}\label{sec:monad}

The \MultiChor API features seven primitive monadic operations,
from which a collection of utilities can be built.
Additionally, there's a suite of pure operations for handing membership and subset proofs,
and some limited operations which can manipulate \inlinecode{Located}
and \inlinecode{Faceted} values.

\paragraph{Pure operations for proofs}
\MultiChor's API uses proof objects both to identify parties and sets of parties,
and as assurance that those parties will be able to perform the described actions at runtime.
In all cases, these are proofs that the identified party is a member of some set
or that the identified set is a subset of some other set.
The suite of operations for building such proofs is founded on four uses of
\inlinecode[text]{gdp}'s \inlinecode{Logic.Proof.axiom}
and instances for \inlinecode{Subset} of \inlinecode{Logic.Classes.Reflexive}
(which affords the proof \inlinecode{refl})
and \inlinecode{Transitive} (which affords the proof \inlinecode{transitive})~\cite{gdp_hackage}.
In situations where the census and all identities are known explicitly,
the proofs \inlinecode{explicitMember} and \inlinecode{explicitSubset} may be used indiscriminately.
(These rely on respective type classes \inlinecode{ExplicitMember}/\inlinecode{ExplicitSubset},
the limitations of which are part of the motivation for using
\inlinecode[text]{gdp}.)
A Template Haskell helper \inlinecode{mkLoc} specializes \inlinecode{explicitMember} for a given 
symbol, allowing easy, clear, and precise reference to monomorphic parties.

Specifying a list of parties via subset proof can be finicky.
To facilitate, the API offers aliases of two of the axioms
than can be used analogously to the empty-list and list-cons constructors:
\begin{minted}[xleftmargin=10pt,fontsize=\small]{haskell}
nobody :: Subset '[] ys
(@@) :: Member x ys -> Subset xs ys
     -> Subset (x ': xs) ys
\end{minted}
For example, in Figure~\ref{fig:card-game} line~18, the first argument to the
\inlinecode{enclaveTo}
says that the census of the enclave is
\inlinecode{"dealer"} and \inlinecode{player} and nobody else.

\paragraph{Pure operations for located values}
\inlinecode{Located} and \inlinecode{Faceted} are both instances of \inlinecode{Wrapped},
indicating that these values can be \inlinecode{un}-wrapped in an appropriate context.
Almost every operation on \inlinecode{Located} and \inlinecode{Faceted} values is dependent
on a relevant census, and so must be performed as part of the \inlinecode{Choreo} monad.
Exceptions pertain to reinterpreting the wrapper-type itself, not the contained data.
Most of these are uninteresting,
but \inlinecode{flatten} is indispensable and must be considered part of the core API.
\begin{minted}[xleftmargin=10pt,fontsize=\small]{haskell}
flatten :: Subset ls ms -> Subset ls ns
        -> Located ms (Located ns a) -> Located ls a
\end{minted}
\inlinecode{flatten} un-nests \inlinecode{Located} layers to parties listed in both layers.
In practice, its primary use is inside the helper-function \inlinecode{enclaveTo};
in Figure~\ref{fig:card-game} line~18 the second argument says
"the second party listed in the census (\inlinecode{player}) knows the return value",
and gets passed to \inlinecode{flatten} so that the result of the enclaved choreography
isn't doubly-\inlinecode{Located}.
(An alternative API, in which \inlinecode{enclave} didn't add a layer of wrapping but
required it's return to be located, would be less expressive.)

\paragraph{Monadic primitives}
The \inlinecode{Choreo} monad is implemented as a freer monad with an explicit census
and an inner-monad type parameter representing
the local computational model of all parties individually.
The underlying data-type has seven constructors, each exposed as a function;
Figure~\ref{fig:monad-api} shows their types and describes their arguments.
\begin{itemize}[leftmargin=12pt, topsep=2pt]
    \item \inlinecode{parallel} runs a local monadic computation in parallel across a list of parties.
          This computation can depend on the identity of the party in question,
          and can use a provided function to unwrap \inlinecode{Located} and \inlinecode{Faceted}
          values.
          The return type is \inlinecode{Faceted} across the relevant parties.
    \item \inlinecode{congruently} runs a pure computation congruently across a list of parties;
          since the values represented by a \inlinecode{Faceted} aren't congruent,
          \inlinecode{congruently} can't use them.
          Since the returned values are congruent,
          \inlinecode{congruently} returns a \inlinecode{Located} value at all of the listed parties.
    \item \inlinecode{comm} is the communication operator;
          it multicasts from a single sender to a list of recipients.
          The result is \inlinecode{Located}.
          See Section~\ref{sec:helpers} for the more user-friendly \inlinecode{~>}.
    \item \inlinecode{enclave} embeds a choreography with a smaller census inside one whom's
          census is a superset.
          In addition to facilitating code-reuse, this is important when one wants to encode
          at the type level that certain parties are not involved in some sub-operation,
          and it interacts with \inlinecode{naked}.
    \item \inlinecode{naked} unwraps a single value that's known to the entire census.
          Monadic binding of \inlinecode{naked} with a continuation gives the same behavior
          as \HasChor's \inlinecode{cond} operation, except that instead of hiding an implicit
          broadcast it requires that any needed communication have already happened.
          Frequently, \inlinecode{naked} will be called inside \inlinecode{enclave};
          this gives the behavior of \HLS's auto-enclaving "case" expressions.
    \item \inlinecode{fanOut} performs a given action parameterized by a party
          for each party in the given subset of the census.
          Unlike \inlinecode{parallel}, which describes actions in the local monad,
          \inlinecode{fanOut}'s body is a choreography over the complete census.
          Arbitrary choreographic behavior can happen inside \inlinecode{fanOut},
          including branching based on the parameter identity,
          so long as the returned value is known to that party and has the given type.
          The return values form a new \inlinecode{Faceted}.
    \item \inlinecode{fanIn} is similar to \inlinecode{fanOut}, but represents \emph{convergence}.
          The return value must be \inlinecode{Located} at a given set of "recipients"
          that do \emph{not} depend on the parameter identity.
          This doesn't necessarily imply any \inlinecode{~>} operations,
          but the obvious use-case is that each \inlinecode{q} in \inlinecode{qs}
          multicasts to \inlinecode{rs}.
          The likely-different return values are combined as a simple \inlinecode{Located} list;
          any other scheme for combining them can be implemented
          in a subsequent \inlinecode{congruently}.
\end{itemize}

\begin{figure*}[tbhp]
    \begin{mdframed}
\begin{minted}[xleftmargin=10pt,linenos,fontsize=\small]{haskell}
-- | Access to the inner "local" monad.
parallel :: Subset ls ps  -- A set of parties who will all perform the action(s) in parallel.
         -> (forall l. Member l ls -- The "loop variable", a party l in ls.
                    -> Unwrap l    -- A function for unwrapping Located or Faceted values
                                   --  known to l.
                    -> m a)  -- The local action(s).
         -> Choreo ps m (Faceted ls a)

-- | Congrunet computation.
congruently :: Subset ls ps  -- The set of parties who will perform the computation.
            -> (Unwraps ls -> a)  -- The computation does not depend on who's performing it.
                                  -- Unwraps doesn't work on Faceted values, unlike Unwrap.
            -> Choreo ps m (Located ls a)

-- | Communication between a sender and some receivers.
comm :: (Show a, Read a, Wrapped w)  -- Both Located and Faceted are instances of Wrapped.
                                     -- Show and Read are the ad-hoc serializaiton system.
     => Member l ps            -- ^ The sender, who is present in the census.
     -> (Member l ls, w ls a)  -- ^ Proof the sender knows the value, the value.
     -> Subset ls' ps          -- ^ The recipients, who are present in the census.
     -> Choreo ps m (Located ls' a)

-- | Lift a choreography of involving fewer parties into the larger party space.
enclave :: Subset ls ps  -- The sub-choreography's census must be in the outer census.
        -> Choreo ls m a  -- The sub-choreography.
        -> Choreo ps m (Located ls a)

-- | Un-locates a value known to the entire census.
naked :: Subset ps qs  -- Proof that everyone knows the value.
      -> Located qs a  --  The value.
      -> Choreo ps m a

-- | Perform a choreographic action for each of several parties,
--   gathering their return values as a `Faceted`.
fanOut :: (Wrapped w)
       => Subset qs ps  -- The parties to loop over.
       -> (forall q. Member q qs  -- The "loop variable", a party q in qs.
                  -> Choreo ps m (w '[q] a))  -- The "loop body", returns a value at q.
       -> Choreo ps m (Faceted qs a)

-- | Perform a given choreography for each of several parties;
--   the return values are aggregated as a list located at the invariant recipients.
fanIn :: Subset qs ps  -- The parties who are fanning-in, or over whom we're looping.
      -> Subset rs ps  -- The recipients, the parties who know the results.
      -> (forall q. Member q qs  -- The "loop variable", a party q in qs.
                 -> Choreo ps m (Located rs a))  -- The "loop body", returns a value at rs.
      -> Choreo ps m (Located rs [a])
\end{minted}
    \caption{The monadic functions for constructing \inlinecode{Choreo} expressions.}
    \label{fig:monad-api}
    \Description{Forty-seven lines of haskell code, giving type signatures for seven functions.}
    \end{mdframed}
\end{figure*}

\subsection{Derived functions}\label{sec:helpers}
We aim for minimalism in engineering \MultiChor's core API to facilitate reasoning about
(and implementing) the freer monad handlers for the centralized semantics and for EPP.
At the same time, we aim for \MultiChor to be used in the wild,
and based on our experience writing examples
we believe that a suite of helper functions is in order.
Appendix~\ref{sec:extra-helpers} lists all the functions used in examples in this paper.
A few are especially common in practice:
\begin{itemize}[leftmargin=12pt, topsep=2pt]
    \item \inlinecode{~>} wraps \inlinecode{comm}.
          In addition to being a convenient infix operator,
          it uses a dedicated class to accept different structures as its first argument.
          When explicit membership can be inferred by the type system,
          it suffices to identify the sender with a \inlinecode{Member sender census}.
          When all owners of the message are present in the census,
          one can use a \inlinecode{Subset owners census} and a \inlinecode{Member sender owners}.
          Otherwise, the usual proofs of presence and ownership will work.
    \item \inlinecode{~~>} sends the result of a local computation directly to recipients
          without binding it to an intermediary variable.
    \item \inlinecode{broadcast} takes the same argument options as \inlinecode{~>}
          and sends the message to the entire census;
          the received value is unwrapped by \inlinecode{naked}.
    \item \inlinecode{locally} wraps \inlinecode{parallel} for use on a single party;
          since there's only one actor, the return value can be \inlinecode{Located}
          instead of \inlinecode{Faceted}.
          Both \inlinecode{locally} and \inlinecode{parallel} have underscore-prefixed
          variants for when the action-lambdas would ignore their arguments,
          and underscore-suffixed variants for discarding \inlinecode{Located ps ()} results.
    \item \inlinecode{singleton} is a polymorphic proof that a party is a member of a list
          containing just themselves. 
\end{itemize}

\begin{figure*}
\begin{mdframed}
\begin{minted}[xleftmargin=10pt,linenos,fontsize=\small]{haskell}
kvs :: (KnownSymbol client) => ReplicationStrategy ps (CLI m) -> Member client ps -> Choreo ps (CLI m) ()
kvs ReplicationStrategy{setup, primary, handle} client = do
  rigging <- setup
  let go = do request <- (client, readRequest) -~> primary @@ nobody
              response <- handle rigging singleton request
              case response of Stopped -> return ()
                               _ -> do client `_locally_` putOutput "Recieved:" response
                                       go
  go

naryReplicationStrategy :: (KnownSymbol primary, KnownSymbols backups, KnownSymbols ps, MonadIO m)
                        => Member primary ps -> Subset backups ps -> ReplicationStrategy ps m
naryReplicationStrategy primary backups = ReplicationStrategy{
      primary
    , setup = servers `_parallel` newIORef (Map.empty :: State)
    , handle = \stateRef pHas request -> do
          request' <- (primary, (pHas, request)) ~> servers
          localResponse <- servers `parallel` \server un ->
              handleRequest (un server stateRef) (un server request')
          responses <- fanIn servers (primary @@ nobody) \server ->
              (server, servers, localResponse) ~> primary @@ nobody
          response <- (primary @@ nobody) `congruently` \un ->
              case nub (un refl responses) of [r] -> r
                                              rs -> Desynchronization rs
          broadcast (primary, response)   }
  where servers = primary @@ backups

data ReplicationStrategy ps m = forall primary rigging. (KnownSymbol primary) =>
  ReplicationStrategy { primary :: Member primary ps
                      , setup :: Choreo ps m rigging
                      , handle :: forall starts w. (Wrapped w)
                               => rigging -> Member primary starts -> w starts Request
                               -> Choreo ps m Response  }

data Request = Put String String  | Get String  | Stop  deriving (Eq, Ord, Read, Show)

data Response = Found String  | NotFound  | Stopped  | Desynchronization [Response]
                deriving (Eq, Ord, Read, Show)

-- | PUT returns the old stored value; GET returns whatever was stored.
handleRequest :: (MonadIO m) => IORef State -> Request -> m Response
handleRequest stateRef (Put key value) = mlookup key <$> modifyIORef stateRef (Map.insert key value)
handleRequest stateRef (Get key) = mlookup key <$> readIORef stateRef
handleRequest _         Stop = return Stopped

mlookup :: String -> State -> Response
mlookup key = maybe NotFound Found . Map.lookup key

type State = Map String String
\end{minted}
\caption{A system for building key-value-store choreographies,
         including an example backup strategy that's polymorphic on the number of backup servers.}
\label{fig:kvs}
\Description{Forty-nine lines of haskell code defining functions "kvs", "naryReplicationStrategy", "handleRequest", and "mlookup", and associated data types.}
\end{mdframed}
\end{figure*}

\section{Case studies}\label{sec:examples}
The attached artifact contains over twenty example choreographies,
including translations into \MultiChor of all the examples in
the \HasChor git repository~\cite{haschor-repo},
key examples from Bates~\&~Near~\cite{bates2024know},
and the examples in this paper.
We present three use-cases here.
Most of our examples have \inlinecode{MonadIO m => CLI m} as their local monad.
\inlinecode{CLI} is perpendicular to the goals of \MultiChor;
it lets a choreography both run in simultaneous shells with human interaction and
run inside of QuickCheck tests.

\subsection{Key-Value Store}\label{sec:kvs}
Shen~\textit{et~al} present four different "key-value store" choreographies of increasing
sophistication.
We continue directly from their forth version to write an \emph{"N-ary replicated KVS"},
presented in Figure~\ref{fig:kvs}.

Like \HasChor's examples, our key-value-store is polymorphic on the identities of all parties
involved, and higher-order in the sense that it takes as an argument some choreographies
that specify how the store should be replicated.
It also features recursive interaction with the client and the ability to report corruption
(de-synchronization) of servers.
A non-replicated system with a single server
(or a fake replication scheme
where humans interactively report key-value pairs on the command line)
can also fit in this scheme.
Figure~\ref{fig:kvs} showcases an N-ary replication scheme
that abstracts over the number of backup servers;
prior systems cannot represent this abstraction.

The choreography \inlinecode{kvs} on line~2 is short;
it sets up the replication system and then recursively passes requests through it
until a \inlinecode{Stop} request has been handled.
\inlinecode{singleton} on line~5 indicates that
only \inlinecode{primary} owns \inlinecode{request};
the type of \inlinecode{handle} on line~31 leaves open the possibility
that other parties may also know the \inlinecode{Request} by taking an argument
\inlinecode{Member primary starts}.
Note that the existence of any backup servers is entirely hidden from \inlinecode{kvs},
but the possibility of such unseen participants is left open
by the polymorphic census \inlinecode{ps}.
In the \inlinecode{fanIn} on line~20
all the servers \emph{including \inlinecode{primary}} send their responses
to \inlinecode{primary} for collation;
these kinds of self-sends are typical and do not incur communication costs.
The \inlinecode{broadcast} on line~25 is necessary because everyone needs to know
whether to break recursion on line~6 or not.

\subsection{Secure Multiparty Computation: GMW}\label{sec:gmw}

\emph{Secure multiparty computation}~\cite{evans2018pragmatic} (MPC) is a family of techniques that allow a group of parties to jointly compute an agreed-upon function of their distributed data without revealing the data or any intermediate results to the other parties. We consider an MPC protocol named Goldreich-Micali-Widgerson (GMW)~\cite{goldreich2019play} after its authors. The GMW protocol requires the function to be computed to be specified as a binary circuit, and each of the parties who participates in the protocol may provide zero or more inputs to the circuit. At the conclusion of the protocol, all participating parties learn the circuit's output.

The GMW protocol is based on two important building blocks: \emph{additive secret sharing}, a method for encrypting distributed data that still allows computing on it, and \emph{oblivious transfer} (OT)~\cite{naor2001efficient}, a building-block protocol in applied cryptography. The GMW protocol starts by asking each party to secret share its input values for the circuit. Then, the parties iteratively evaluate the gates of the circuit while keeping the intermediate values secret shared. Oblivious transfer is used to evaluate AND gates. When evaluation finishes, the parties reveal their secret shares of the output to decrypt the final result.

\begin{figure*}
\begin{mdframed}
\begin{minted}[xleftmargin=10pt,linenos,fontsize=\small]{haskell}
genShares :: (MonadIO m) => [LocTm] -> Bool -> m [(LocTm, Bool)]
genShares partyNames x = do
  freeShares <- replicateM (length partyNames - 1) $ liftIO randomIO -- generate n-1 random shares
  return $ zip partyNames $ xor (x : freeShares) : freeShares        -- make the sum equal to x

secretShare :: forall parties p m. (KnownSymbols parties, KnownSymbol p, MonadIO m)
            => Member p parties -> Located '[p] Bool -> Choreo parties m (Faceted parties Bool)
secretShare p value = do
  shares <- p `locally` \un -> genShares (toLocs $ allOf @parties) (un singleton value)
  allOf @parties `fanOut` \q ->
    (p, \un -> return $ fromJust $ toLocTm q `lookup` un singleton shares) ~~> q @@ nobody

reveal :: forall ps m. (KnownSymbols ps) => Faceted ps Bool -> Choreo ps m Bool
reveal shares = let ps = allOf @ps
                in xor <$> ((fanIn ps ps \p -> (p, (p, shares)) ~> ps) >>= naked ps)
\end{minted}
\caption{Choreographies for secret sharing \inlinecode{p}'s secret value among \inlinecode{parties}
         and for revealing a secret-shared value.
         \inlinecode{p} constructs secret shares locally (line~9),
         then sends one share to each party (lines~10--11), including themselves.
         The result is a \inlinecode{Faceted} value---each
         party has one secret share of the secret.}
\label{fig:secret-sharing}
\Description{Fifteen lines of Haskell defining functions "genShares", "secretShare", and "reveal".}
\end{mdframed}
\end{figure*}

\paragraph{Additive secret sharing}
We begin by describing additive secret sharing, an extremely common building block in MPC protocols. A secret bit $x$ can be \emph{secret shared} by generating $n$ random \emph{shares} $s_1, \dots, s_n$ such that $x = \sum_{i=1}^n s_i$. If $n-1$ of the shares are generated uniformly and independently randomly, and the final share is chosen to satisfy the property above, then the shares can be safely distributed to the $n$ parties without revealing $x$---recovering $x$ requires access to all $n$ shares. Importantly, secret shares are \emph{additively homomorphic}---adding together shares of secrets $x$ and $y$ produces a share of $x+y$. 

\MultiChor choreograpies for performing secret sharing in the arithmetic field of booleans appear in Figure~\ref{fig:secret-sharing}. The function \inlinecode{secretShare} takes a single secret bit located at party \inlinecode{p} and returns a \inlinecode{Faceted} bit located at all parties, such that the bits held by the parties sum up to the original secret.
\inlinecode{reveal} takes such a shared value and broadcasts
all the shares so everyone can reconstruct the plain-text.

\paragraph{Oblivious transfer}
The other important building block of the GMW protocol is oblivious transfer (OT)~\cite{naor2001efficient}. OT is a 2-party protocol between a \emph{sender} and a \emph{receiver}. In the simplest variant (\emph{1 out of 2} OT, used in GMW), the sender inputs two secret bits $b_1$ and $b_2$, and the receiver inputs a single secret \emph{select bit} $s$. If $s=0$, then the receiver receives $b_1$. If $s=1$, then the receiver receives $b_2$. Importantly, the sender does \emph{not} learn which of $b_1$ or $b_2$ has been selected, and the receiver does \emph{not} learn the non-selected value.

The type for the  \MultiChor choreography for 1-of-2 oblivious transfer is:
\begin{minted}[xleftmargin=10pt,fontsize=\small]{haskell}
ot2 :: ... => Located '[sender] (Bool, Bool)
           -> Located '[receiver] Bool
           -> Choreo '[sender, receiver] (CLI m) 
                (Located '[receiver] Bool)
\end{minted}
The complete definition appears in Figure~\ref{fig:oblivious-transfer} in Appendix~\ref{sec:gmw_appendix}. Importantly, oblivious transfer is a \emph{two-party protocol}, it would be a type-error for any third-parties to be involved in the implementation. \MultiChor's \inlinecode{Faceted} values and utilities for type-safe embedding of enclaved sub-protocols within arbitrarily large censuses make it possible to embed the use of pairwise oblivious transfer between parties in a general version of multi-party GMW.

\begin{figure*}
\begin{mdframed}
\begin{minted}[xleftmargin=10pt,linenos,fontsize=\small]{haskell}
gmw :: forall parties m. (KnownSymbols parties, MonadIO m, CRT.MonadRandom m)
    => Circuit parties -> Choreo parties (CLI m) (Faceted parties Bool)
gmw circuit = case circuit of
  InputWire p -> do        -- process a secret input value from party p
    value :: Located '[p] Bool <- p `_locally` getInput "Enter a secret input value:"
    secretShare p value
  LitWire b -> do          -- process a publicly-known literal value
    let partyNames = toLocs (allOf @parties)
        shares = partyNames `zip` (b : repeat False)
    allOf @parties `fanOut` \p -> p `_locally` pure (fromJust $ toLocTm p `lookup` shares)
  AndGate l r -> do        -- process an AND gate
    lResult <- gmw l; rResult <- gmw r;
    fAnd lResult rResult
  XorGate l r -> do        -- process an XOR gate
    lResult <- gmw l; rResult <- gmw r
    allOf @parties `parallel` \p un -> pure $ xor [un p lResult, un p rResult]
\end{minted}
\caption{A choreography for the GMW protocol. The choreography works for an arbitrary number of parties, and leverages the \inlinecode{fAnd} choreography defined in Figure~\ref{fig:fand} to compute the results of AND gates.}
\label{fig:gmw}
\Description{Sixteen lines of haskell code defining a choreographic function "gmw".}
\end{mdframed}
\end{figure*}

\begin{figure*}
\begin{mdframed}
\begin{minted}[xleftmargin=10pt,linenos,fontsize=\small]{haskell}
fAnd :: forall parties m. (KnownSymbols parties, MonadIO m, CRT.MonadRandom m)
     => Faceted parties Bool -> Faceted parties Bool -> Choreo parties (CLI m) (Faceted parties Bool)
fAnd uShares vShares = do
  let partyNames = toLocs (allOf @parties)
      genBools = mapM (\name -> (name,) <$> randomIO)
  a_j_s :: Faceted parties [(LocTm, Bool)] <- allOf @parties `_parallel` genBools partyNames
  bs :: Faceted parties Bool  <- allOf @parties `fanOut` \p_j -> do
      let p_j_name = toLocTm p_j
      b_i_s <- fanIn (allOf @parties) (p_j @@ nobody) \p_i ->
        if toLocTm p_i == p_j_name
          then p_j `_locally` pure False
          else do
              bb <- p_i `locally` \un -> let a_ij = fromJust $ lookup p_j_name (un p_i a_j_s)
                                             u_i = un p_i uShares
                                         in pure (xor [u_i, a_ij], a_ij)
              enclaveTo (p_i @@ p_j @@ nobody) (listedSecond @@ nobody) (ot2 bb $ localize p_j vShares)
      p_j `locally` \un -> pure $ xor $ un singleton b_i_s
  allOf @parties `parallel` \p_i un ->
    let name = toLocTm p_i
        ok p_j = p_j /= name
        computeShare u v a_js b = xor $ [u && v, b] ++ [a_j | (p_j, a_j) <- a_js, ok p_j]
    in pure $ computeShare (un p_i uShares) (un p_i vShares) (un p_i a_j_s) (un p_i bs)
\end{minted}
\caption{A choreography for computing the result of an AND gate on secret-shared inputs using pairwise oblivious transfer. The choreography works for an arbitrary number of parties, and leverages the 1 out of 2 OT defined earlier.}
\label{fig:fand}
\Description{Twenty two lines of haskell code defining a choreographic function "fAnd".}
\end{mdframed}
\end{figure*}

\paragraph{The GMW protocol}
The complete GMW protocol operates as summarized earlier, by secret sharing input values and then evaluating the circuit gate-by-gate. Our implementation as a \MultiChor choreography appears in Figure~\ref{fig:gmw}, defined as a recursive function over the structure of the circuit. The choreography returns a \inlinecode{Faceted} value, representing the secret-shared output of the circuit. For ``input'' gates (lines~4--6), the choreography runs the secret sharing protocol in Figure~\ref{fig:secret-sharing} to distribute shares of the secret value. For XOR gates (lines~14--16), the parties recursively run the GMW protocol to compute the two inputs to the gate and then each party computes one share of the gate's output by XORing their shares of the inputs. This approach leverages the additive homomorphism of additive secret shares. For AND gates (lines~11--13), the parties compute shares of the gate's inputs, then use the \inlinecode{fAnd} protocol to perform multiplication of the two inputs. Since additive secret shares are not multiplicatively homomorphic, this operation leverages the oblivious transfer protocol to perform the multiplication.

\paragraph{Computing secret-shared AND via OT}
To compute the result of an AND gate, the parties compute \emph{pair-wise} ANDs using their respective shares of the input values, then use the results to derive shares of the gate's output. The \inlinecode{fAnd} choreography (Figure~\ref{fig:fand}) takes \inlinecode{Faceted} values holding the parties' shares of the input values, and returns a \inlinecode{Faceted} value representing each party's share of the output. On line~7, the parties perform a \inlinecode{fanOut} to begin the pairwise computation; the \inlinecode{fanIn} on line~9 completes the pairing for each computation, and uses \inlinecode{enclaveTo} (line 16) to embed pairwise OTs (via \inlinecode{ot2}) in the larger set of parties.

Our implementation of GMW leverages \MultiChor's
\\
\inlinecode{Faceted} values and utilities for type-safe parallel, enclaved, and pairwise choreographies to build a fully-general implementation of the protocol that works for an arbitrary number of parties.

\begin{figure*}
\begin{mdframed}
\begin{minted}[xleftmargin=10pt,linenos,fontsize=\small]{haskell}
lottery
  :: forall clients servers analyst census m _serv1 _serv2 _servTail _client1 _client2 _clientTail
   . ( KnownSymbols clients, KnownSymbols servers, KnownSymbol analyst, MonadIO m
     , (_serv1 ': _serv2 ': _servTail) ~ servers -- There must be at least be two servers
     , (_client1 ': _client2 ': _clientTail) ~ clients -- There must be at least be two clients
     )
  => Subset clients census -> Subset servers census -> Member analyst census
  -> Choreo census (CLI m) ()
lottery clients servers analyst = do
  secret <- _parallel clients (getInput @Fp "secret:")
  clientShares <- clients `parallel` \client un -> do
      freeShares <- liftIO $ replicateM (length serverNames - 1) $ randomIO @Fp
      return $ serverNames `zip` (un client secret - sum freeShares : freeShares)
  serverShares <- fanOut servers (\server ->
      fanIn clients (inSuper servers server @@ nobody) (\client -> do
          serverShare <- inSuper clients client `locally` \un ->
                           pure $ fromJust $ lookup (toLocTm server) $ un client clientShares
          (inSuper clients client, serverShare) ~> inSuper servers server @@ nobody
        )
    )
  -- 1) Each server selects a random number; τ is some multiple of the number of clients.
  ρ <- _parallel servers (getInput "A random number from 1 to τ:")
  -- Salt value
  ψ <- _parallel servers (randomRIO (2^(18::Int), 2^(20::Int)))
  -- 2) Each server computes and publishes the hash α = H(ρ, ψ) to serve as a commitment
  α <- parallel servers \server un -> pure $ hash (un server ψ) (un server ρ)
  α' <- fanIn servers servers ( \server -> (server, servers, α) ~> servers )
  -- 3) Every server opens their commitments by publishing their ψ and ρ to each other
  ψ' <- fanIn servers servers ( \server -> (server, servers, ψ) ~> servers )
  ρ' <- fanIn servers servers ( \server -> (server, servers, ρ) ~> servers )
  -- 4) All servers verify each other's commitment by checking α = H(ρ, ψ)
  parallel_ servers (\server un ->
      unless (un server α' == zipWith hash (un server ψ') (un server ρ'))
             (liftIO $ throwIO CommitmentCheckFailed)
    )
  -- 5) If all the checks are successful, then sum random values to get the random index.
  ω <- servers `congruently` (\un -> sum (un refl ρ') `mod` length (toLocs clients))
  chosenShares <- servers `parallel` (\server un -> pure $ un server serverShares !! un server ω)
  -- Servers forward shares to an analyist.
  allShares <- fanIn servers (analyst @@ nobody) (\server ->
      (server, servers, chosenShares) ~> analyst @@ nobody
    )
  analyst `locally_` \un -> putOutput "The answer is:" $ sum $ un singleton allShares
 where serverNames = toLocs servers
       hash :: Int -> Int -> Digest Crypto.SHA256
       hash ρ ψ = Crypto.hash $ toStrict (Binary.encode ρ <> Binary.encode ψ)
\end{minted}
\caption{A federated-lottery protocol.
  One of the secret values chosen by the clients is revealed to the analyst;
  as long as at least one server acts honestly
  (randomly generates their \inlinecode{ρ} on line 22),
  the choice of which value to reveal will be random.
  Only the analyst learns any of the clients' secrets;
  they only learn the one secret, and they do not learn which one it was.
  The algorithm-step numbers and the unicode variable names align with
  the instructions in Section 6.2 of Keller~\textit{et~al}~\cite{dprio2023}.
  Client secrets are chosen at-will from a finite field (the type \inlinecode{Fp});
  we used the finite field of size 999983.}
\label{fig:lottery}
\Description{Forty-six lines of haskell code defining a choreography named "lottery".}
\end{mdframed}
\end{figure*}

\subsection{Federated lottery}
DPrio~\cite{dprio2023} is a recent variation on a secure-aggregation
protocol called Prio~\cite{corrigan2017prio}.
DPrio supports all of the same security guarantees as Prio and adds a layer of
differential privacy (DP)
so that client inputs cannot be inferred by the receiving analyst.
To avoid summarizing the polynomial-based zero-knowlege-proofs used in Prio,
we excerpt part of DPrio as a stand-alone protocol for this case study.
The key mechanism of DPrio is that all clients generate noise for DP and
the servers randomly choose whose noise to actually use.
The choreography in Figure~\ref{fig:lottery} implements this process,
except that instead of adding a selected "noise" to their shares of an earlier aggregation,
a value selected from those submitted by the clients is forwarded directly to the analyst
to be revealed.

In Figure~\ref{fig:lottery} lines 10--20 the clients choose their inputs and secret-share them to
the servers.
This process is basically the same as in GMW, except that the field is larger
and the parties generating and receiving shares are distinct subsets of the census.
Lines 21--38 implement the lottery itself.
Finally, the analyst receives shares of the chosen client's secret and sums them together to
learn the final result on line 43.

A more casual lottery could be implemented by having one server choose a client-index at random
and inform the other servers of the choice,
but then everyone would have to trust that the choice truly was random.
DPrio ensures the fairness of the lottery in two steps:
First, \emph{all} the servers independently generate random values up to some multiple
of the number of clients.
Second, before the clients reveal their randomness to each other,
they \emph{commit} to their randomness by broadcasting salted hashes.
The actual client-index used is the modulo of the sum of these random values.
The commitment process just prevents any server from waiting until the end to select their value;
without it the last server would be able to calculate their "random" value to result in
any index they wanted.
We represent the result of a failed commitment check with an \inlinecode{IO} error
at any parties that detect it, which will prevent the choreography from completing.

The three \inlinecode{parallel} blocks on lines 22, 24, and 26 of Figure~\ref{fig:lottery}
could be combined into one without changing the semantics of the choreography,
but a \inlinecode{Faceted} tuple can't be unpacked by pattern matching,
so using \inlinecode{ρ}, \inlinecode{ψ}, and \inlinecode{α}
would become more complicated.
In contrast, it would not be safe to combine the \inlinecode{fanIn} on line 27
with the ones on lines 29 and 30;
it's precisely this sequential separation that ensures no server sends their \inlinecode{ρ}
until they've received all the \inlinecode{α'}.

\section{Conclusions}\label{sec:conclusion}

We have presented \MultiChor, an eDSL library for choreographic programming with
multiply located values, multicast, enclaving, and census polymorphism.
\MultiChor has type-safe KoC management
while avoiding performance penalties due to extra communication.
\MultiChor uses proof objects
both to ensure that choreographies are projectable and as identifiers
for communicating parties.
\MultiChor offers new operations such as \inlinecode{congruently},
\inlinecode{fanOut}, and \inlinecode{fanIn}.
We've presented several example choreographies to show how \MultiChor can implement
real distributed computations including the GMW protocol for secure multi-party computation.
We believe \MultiChor is expressive and an appropriate tool for writing
type-safe choreographic programs in the wild.

\begin{acks}

\MultiChor builds on top of the core engineering of \HasChor,
and we thank Gan Shen for permission to use his copyright material.

This material is based upon work supported by the National Science Foundation under Grant No. 2238442 and by the Cold Regions Research and Engineering Laboratory (ERDC-CRREL)
under Contract No. W913E521C0003. Any opinions, findings and conclusions or recommendations expressed in this material are those of the author(s) and do not necessarily reflect the views of the National Science Foundation or the Cold Regions Research and Engineering Laboratory.
\end{acks}

\bibliographystyle{ACM-Reference-Format}
\bibliography{refs}

\clearpage
\appendix

\section{Additional functions}\label{sec:extra-helpers}
Figure~\ref{fig:monad-api} showed the foundational monadic functions
\inlinecode{parallel}, \inlinecode{congruently},
\inlinecode{comm},
\inlinecode{enclave}, \inlinecode{naked}, \inlinecode{fanOut},
and \inlinecode{fanIn}.
Figure~\ref{fig:other-api} shows type signatures for the entire
library (as exported), in alphabetical order.
Some functions are "core" in the sense that they can't be written
without reference to \MultiChor's internal machinery.
Other functions are "helper functions",
which are implemented entirely using the exposed core functions.

\begin{figure*}
\begin{mdframed}
\begin{minted}[xleftmargin=10pt,linenos,fontsize=\small]{haskell}
(-~>) :: forall a l ls' m ps. (Show a, Read a, KnownSymbol l, KnownSymbols ls')
      => (Member l ps, m a) -> Subset ls' ps -> Choreo ps m (Located ls' a)
infix 4 -~>

(~~>) :: forall a l ls' m ps. (Show a, Read a, KnownSymbol l, KnownSymbols ls')
      => (Member l ps, Unwrap l -> m a) -> Subset ls' ps -> Choreo ps m (Located ls' a)
infix 4 ~~>

(@@) :: Member x ys -> Subset xs ys -> Subset (x ': xs) ys
infixr 5 @@

(~>) :: (Show a, Read a, KnownSymbol l, KnownSymbols ls', CanSend s l a ls ps w)
     => s  -- ^ The message argument can take three forms:
           --     `(Member sender census, wrapped owners a)` where
           --         the sender is explicitly listed in owners,
           --     `(Member sender owners, Subset owners census, wrapped owners a)`, or
           --     `(Member sender census, (Member sender owners, wrapped owners a)`.
     -> Subset ls' ps -> Choreo ps m (Located ls' a)
infix 4 ~>

allOf :: forall ps. Subset ps ps

Backend :: Type -> Constraint  -- A phonebook for running Network monad expressions.

broadcast :: forall l a ps ls w m s.
             (Show a, Read a, KnownSymbol l, KnownSymbols ps, CanSend s l a ls ps w)
          => s -> Choreo ps m a

Choreo :: [LocTy] -> (Type -> Type) -> Type -> Type
type Choreo ps m = Freer (ChoreoSig ps m)

comm :: (Show a, Read a, KnownSymbol l, KnownSymbols ls', Wrapped w)
     => Member l ps -> (Member l ls, w ls a) -> Subset ls' ps
     -> Choreo ps m (Located ls' a)
infix 4 `comm`

cond :: (KnownSymbols ls)
     => (Subset ls ps, (Subset ls qs, Located qs a)) -> (a -> Choreo ls m b)
     -> Choreo ps m (Located ls b)

congruently :: (KnownSymbols ls)
              => Subset ls ps -> (Unwraps ls -> a) -> Choreo ps m (Located ls a)
infix 4 `congruently`

consSet :: Subset xs (x ': xs)

consSub :: Subset xs ys -> Member x ys -> Subset (x ': xs) ys

consSuper :: forall xs ys y. Subset xs ys -> Subset xs (y ': ys)
\end{minted}
\caption{The \MultiChor API, part 1/4.}
\label{fig:other-api}
\Description{A detailed API description, across four pages, in Haskell.}
\end{mdframed}
\end{figure*}
\begin{figure*}\ContinuedFloat
\begin{mdframed}
\begin{minted}[xleftmargin=10pt,linenos,firstnumber=50,fontsize=\small]{haskell}
enclave :: (KnownSymbols ls)
        => Subset ls ps -> Choreo ls m a -> Choreo ps m (Located ls a)
infix 4 `enclave`

enclaveTo :: forall ls a rs ps m. (KnownSymbols ls)
          => Subset ls ps -> Subset rs ls -> Choreo ls m (Located rs a)
          -> Choreo ps m (Located rs a)
infix 4 `enclaveTo`

enclaveToAll :: forall ls a ps m. (KnownSymbols ls)
             => Subset ls ps -> Choreo ls m (Located ls a) -> Choreo ps m (Located ls a)
infix 4 `enclaveToAll`

epp :: (Monad m) => Choreo ps m a -> LocTm -> Network m a

ExplicitMember :: forall k. k -> [k] -> Constraint
explicitMember :: forall k (x :: k) (xs :: [k]). ExplicitMember x xs => Member x xs

ExplicitSubset :: forall {k}. [k] -> [k] -> Constraint
explicitSubset :: forall {k} (xs :: [k]) (ys :: [k]). ExplicitSubset xs ys => Subset xs ys

Faceted :: [LocTy] -> Type -> Type

fanIn :: (KnownSymbols qs, KnownSymbols rs)
       => Subset qs ps -> Subset rs ps
       -> (forall q. (KnownSymbol q)
           => Member q qs -> Choreo ps m (Located rs a))
       -> Choreo ps m (Located rs [a])

fanOut :: (KnownSymbols qs, Wrapped w)
       => Subset qs ps
       -> (forall q. (KnownSymbol q) => Member q qs -> Choreo ps m (w '[q] a))
       -> Choreo ps m (Faceted qs a)

flatten :: Subset ls ms -> Subset ls ns -> Located ms (Located ns a)
        -> Located ls a
infix 3 `flatten`

fracture :: forall ls a. (KnownSymbols ls) => Located ls a -> Faceted ls a

inSuper :: Subset xs ys -> Member x xs -> Member x ys

IsMember :: forall k. k -> [k] -> Type

IsSubset :: forall k. [k] -> [k] -> Type

KnownSymbols :: [Symbol] -> Constraint  -- lift KnownSymbol to type-level lists

listedFifth :: forall p5 p4 p3 p2 p1 ps. Member p5 (p1 ': p2 ': p3 ': p4 ': p5 ': ps)

listedFirst :: forall p1 ps. Member p1 (p1 ': ps)

listedForth :: forall p4 p3 p2 p1 ps. Member p4 (p1 ': p2 ': p3 ': p4 ': ps)

listedSecond :: forall p2 p1 ps. Member p2 (p1 ': p2 ': ps)
\end{minted}
\caption{The \MultiChor API, part 2/4.}
\Description{A detailed API description, across four pages, in Haskell.}
\end{mdframed}
\end{figure*}
\begin{figure*}\ContinuedFloat
\begin{mdframed}
\begin{minted}[xleftmargin=10pt,linenos,firstnumber=105,fontsize=\small]{haskell}
listedSixth :: forall p6 p5 p4 p3 p2 p1 ps. Member p6 (p1 ': p2 ': p3 ': p4 ': p5 ': p6 ': ps)

listedThird :: forall p3 p2 p1 ps. Member p3 (p1 ': p2 ': p3 ': ps)

localize :: (KnownSymbol l) => Member l ls -> Faceted ls a -> Located '[l] a

locally :: (KnownSymbol (l :: LocTy))
        => Member l ps -> (Unwrap l -> m a) -> Choreo ps m (Located '[l] a)
infix 4 `locally`

locally_ :: (KnownSymbol l) => Member l ps -> (Unwrap l -> m ()) -> Choreo ps m ()
infix 4 `locally_`

_locally :: (KnownSymbol l) => Member l ps -> m a -> Choreo ps m (Located '[l] a)
infix 4 `_locally`

_locally_ :: (KnownSymbol l) => Member l ps -> m () -> Choreo ps m ()
infix 4 `_locally_`

Located :: [LocTy] -> Type -> Type

LocTm :: Type
type LocTm = String

LocTy :: Type
type LocTy = Symbol

Member :: forall {k}. k -> [k] -> Type
type Member x xs = Proof (IsMember x xs)

mkLoc :: String -> Q [Dec]  -- Template Haskell

naked :: Subset ps qs -> Located qs a -> Choreo ps m a
infix 4 `naked`

Network :: (Type -> Type) -> Type -> Type

NetworkSig :: (Type -> Type) -> Type -> Type

nobody :: Subset '[] ys

parallel :: (KnownSymbols ls)
         => Subset ls ps -> (forall l. (KnownSymbol l) => Member l ls -> Unwrap l -> m a)
         -> Choreo ps m (Faceted ls a)

parallel_ :: forall ls ps m. (KnownSymbols ls)
          => Subset ls ps -> (forall l. (KnownSymbol l) => Member l ls -> Unwrap l ->m ())
          -> Choreo ps m ()

_parallel :: forall ls a ps m. (KnownSymbols ls)
          => Subset ls ps -> m a -> Choreo ps m (Faceted ls a)

recv :: forall a (m :: Type -> Type). Read a => LocTm -> Network m a

run :: forall (m :: Type -> Type) a. m a -> Network m a
\end{minted}
\caption{The \MultiChor API, part 3/4.}
\Description{A detailed API description, across four pages, in Haskell.}
\end{mdframed}
\end{figure*}
\begin{figure*}\ContinuedFloat
\begin{mdframed}
\begin{minted}[xleftmargin=10pt,linenos,firstnumber=160,fontsize=\small]{haskell}
runChoreo :: forall ps b m. Monad m => Choreo ps m b -> m b

runNetwork :: (Backend c, MonadIO m) => c -> LocTm -> Network m a -> m a

send :: forall a (m :: Type -> Type). Show a => a -> [LocTm] -> Network m ()

singleton :: forall p. Member p (p ': '[])

Subset :: forall {k}. [k] -> [k] -> Type
type Subset xs ys = Proof (IsSubset xs ys)

toLocs :: forall (ls :: [LocTy]) (ps :: [LocTy]). KnownSymbols ls => Subset ls ps -> [LocTm]

toLocTm :: forall (l :: LocTy) (ps :: [LocTy]). KnownSymbol l => Member l ps -> LocTm

Unwrap :: LocTy -> Type
type Unwrap (l :: LocTy) = forall ls a w. (Wrapped w) => Member l ls -> w ls a -> a

Unwraps :: [LocTy] -> Type
type Unwraps (qs :: [LocTy]) = forall ls a. Subset qs ls -> Located ls a -> a

Wrapped :: ([Symbol] -> Type -> Type) -> Constraint
\end{minted}
\caption{The \MultiChor API, part 4/4.}
\Description{A detailed API description, across four pages, in Haskell.}
\end{mdframed}
\end{figure*}

\section{Additional Choreographies for the GMW Protocol}
\label{sec:gmw_appendix}

\paragraph{Oblivious transfer.}
Our implementation of oblivious transfer leverages a common strategy for building OT from public-key encryption. First, the receiver generates two public keys and one secret key (line~23); one of the public keys is real, and corresponds to the secret key, while the other is chosen at random from the space of public keys and has no corresponding secret key. The select bit determines the ordering of the public keys. The receiver sends the public keys to the sender (lines~24--25); the sender encrypts both $b_1$ and $b_2$ with the corresponding public keys and sends the ciphertexts back to the receiver (lines~26--27). Finally, the receiver decrypts the selected value (lines~28--30).

The sender treats both $b_1$ and $b_2$ in the same way, and cannot tell which public key is real and which one is fake---so the sender does not learn which value was selected. The receiver gets both encrypted values, but can only decrypt one of them, since only one of the public keys used has a corresponding secret key (the other public key is totally random). This version of OT is secure only against \emph{honest but curious} or \emph{passive} adversaries, who observe network communications but do not change the behavior of the parties, since an actively malicious receiver could create two actual key pairs in the first step of the protocol and thus be able to decrypt both $b_1$ and $b_2$ at the end. More complicated variants of the protocol can defend against this kind of attack.

\begin{figure*}
\begin{mdframed}
\begin{minted}[xleftmargin=10pt,linenos,fontsize=\small]{haskell}
genKeys :: (CRT.MonadRandom m) => Bool -> m (RSA.PublicKey, RSA.PublicKey, RSA.PrivateKey)
genKeys s = do -- Generate keys for OT. One key is real, and one is fake - select bit decides
  (pk, sk) <- genKeyPair
  fakePk <- generateFakePK
  return $ if s then (pk, fakePk, sk) else (fakePk, pk, sk)

encryptS :: (CRT.MonadRandom m) => -- Encryption based on select bit
            (RSA.PublicKey, RSA.PublicKey) -> Bool -> Bool -> m (ByteString, ByteString)
encryptS (pk1, pk2) b1 b2 = do c1 <- encryptRSA pk1 b1; c2 <- encryptRSA pk2 b2; return (c1, c2)

decryptS :: (CRT.MonadRandom m) => -- Decryption based on select bit
       (RSA.PublicKey, RSA.PublicKey, RSA.PrivateKey) -> Bool -> (ByteString, ByteString) -> m Bool
decryptS (_, _, sk) s (c1, c2) = if s then decryptRSA sk c1 else decryptRSA sk c2

-- One out of two OT
ot2 :: (KnownSymbol sender, KnownSymbol receiver, MonadIO m, CRT.MonadRandom m) =>
  Located '[sender] (Bool, Bool) -> Located '[receiver] Bool
  -> Choreo '[sender, receiver] (CLI m) (Located '[receiver] Bool)
ot2 bb s = do
  let sender = listedFirst :: Member sender '[sender, receiver]
  let receiver = listedSecond :: Member receiver '[sender, receiver]

  keys <- receiver `locally` \un -> liftIO $ genKeys $ un singleton s
  pks <- (receiver, \un -> let (pk1, pk2, _) = un singleton keys
                           in return (pk1, pk2)) ~~> sender @@ nobody
  encrypted <- (sender, \un -> let (b1, b2) = un singleton bb
                               in liftIO $ encryptS (un singleton pks) b1 b2) ~~> receiver @@ nobody
  receiver `locally` \un -> liftIO $ decryptS (un singleton keys)
                                                           (un singleton s)
                                                           (un singleton encrypted)
\end{minted}
\caption{A choreography for performing 1 out of 2 oblivious transfer (OT) using RSA public-key encryption. The choreography involves exactly two parties, \inlinecode{sender} and \inlinecode{receiver}.}
\label{fig:oblivious-transfer}
\Description{Thirty lines of haskell defining functions "genKeys", "encryptS", "decryptS", and the choreography "ot2".}
\end{mdframed}
\end{figure*}

\end{document}